\documentclass[review]{elsarticle}

\usepackage{lineno,hyperref}
\modulolinenumbers[5]

\journal{Journal of \LaTeX\ Templates}

\begin{document}

\begin{frontmatter}

\title{Coexistence of antiferromagnetism and superconductivity of t-t$^\prime$-J model on honeycomb lattice}

\author[Yin Zhong]{Yin Zhong\corref{mycorrespondingauthor}}
\address{Center for Interdisciplinary Studies $\&$ Key Laboratory for
Magnetism and Magnetic Materials of the MoE, Lanzhou University, Lanzhou 730000, China}

\cortext[mycorrespondingauthor]{Corresponding author}
\ead{zhongy05@hotmail.com (telephone number:+008615193133526)}
\author{Lan Zhang}
\address{Center for Interdisciplinary Studies $\&$ Key Laboratory for
Magnetism and Magnetic Materials of the MoE, Lanzhou University, Lanzhou 730000, China}
\author{Han-Tao Lu}
\address{Center for Interdisciplinary Studies $\&$ Key Laboratory for
Magnetism and Magnetic Materials of the MoE, Lanzhou University, Lanzhou 730000, China}
\author{Hong-Gang Luo}
\address{Center for Interdisciplinary Studies $\&$ Key Laboratory for
Magnetism and Magnetic Materials of the MoE, Lanzhou University, Lanzhou 730000, China}
\address{Beijing Computational Science Research Center, Beijing 100084, China}

\begin{abstract}
Motivated by recent experimental study of antiferromagnetic property of honeycomb compound In$_{3}$Cu$_{2}$VO$_{9}$ [Yan \textit{et al.}, PRB \textbf{85}, 085102 (2012)], we explore possible superconductivity and its coexistence with collinear antiferromagnetism. Explicitly, we use the t-t$^\prime$-J model on the honeycomb lattice as our starting point and employ the slave-boson mean-field theory. In the antiferromagnetic normal state, the characteristic doping evolution of Fermi surface shows that only one effective singe band is active, which suggests that the potential pairing symmetry is more likely the time-reversal symmetry breaking $d+id$, rather than the extended $s$-wave pairing structure. It is found that this superconducting state coexists with the antiferromagnetism in a broad doping regime, which is consistent with the numerical calculations. The local density of states and its thermodynamic property of the superconducting state has been studied in detail with an effective single-band picture for understanding other physical observable such as superfluid density. The present work may be useful in experimentally exploring possible superconductivity of this kind of materials on the honeycomb lattice and contributes to the understanding of the unconventional superconductivity on general two-dimensional correlated electron systems.
\end{abstract}


\end{frontmatter}

\linenumbers

\section{Introduction} \label{intr}
The t-J-like model is one of the most fundamental theoretical model in condensed matter physics, which is believed to be able to capture the essence of high temperature superconductivity
of cuprate.\cite{Zhang1988} However, in spite of twenty years' extensive and intensive study, the basic feature of such model is still controversial due to its nonperturbative feature,\cite{Anderson2004,Lee2006} which hinders the further understanding of the unconventional superconductivity.

Recently, the t-J model on the honeycomb lattice has re-attracted much attention and has been investigated by the renormalized mean-field theory (RMFT)\cite{Anderson2004, Wu2013} and the Grassmann tensor product state (GTPS) approach. \cite{Gu2013} Theoretically, this was motivated by comparing it to the more standard and more familiar models on the square lattice. It is expected that such parallel study may provide more insight into the secret of cuprates. On the other hand, the experimentally found insulating compound In$_{3}$Cu$_{2}$VO$_{9}$ has a honeycomb lattice structure \cite{Mydosh2008} and its magnetic property has been further explored by Yan \textit{et al.} \cite{Chen2012}. The basic electronic band structure of such compound is calculated by the density functional theory.\cite{Liu2013} In the undoped case, the ground state of this material is probably a N\'eel antiferromagnet and the Co$^{2+}$ replacement of Cu$^{2+}$ will lead to an antiferromagnetic long-range order. If doping, it may allow one to introduce the t-J-like model to understand the magnetic property and possible superconductivity of this material, similar to the case in cuprates. [Doped graphene may also be relevant to t-J-like model if strong coupling condition is achieved.\cite{Baskaran2002,Uchoa2007,Schaffer2007,Honerkamp2008,Pathak2010}]

Actually, the numerical GTPS simulation found a large coexistent regime for the superconducting (SC) and antiferromagnetic (AF) states, \cite{Gu2013} which has been completely missed by the RMFT calculation. \cite{Wu2013} Even the AF state itself has not
been captured by their mean-field theory. However, based on our knowledge on electron-doped cuprates,\cite{Kusko2002,Yuan2004,Xiang2009,Armitage2010} the slave-boson mean-field (SBMF) theory can correctly reproduce the experimentally observed doping dependent phase diagram of t-J-like model, particularly when the AF long-ranged order exists. Therefore, in the present paper, we try to understand the doping dependent physics of the t-t$^\prime$-J model on the honeycomb lattice by using the simple but reliable SBMF theory, which is able to provide more physical picture of the t-J-like model, when comparing to the numerically sophisticated GTPS method.

We first focus on the electronic structure of the antiferromagnetic normal state. The characteristic doping evolution of magnetization and Fermi surface indicate that only one effective singe band is active and is responsible for possible pairing instability. Due to such single-band feature, the candidate pairing state should be the time-reversal symmetry breaking $d+id$ structure, rather than the extended $s$-wave. With this pairing symmetry, we find that the superconductivity and antiferromagnetism can coexist in a broad doping regime, which is quantitatively consistent with the numerical calculation.\cite{Gu2013} To further explore the property of the superconducting states, we discuss the local density of states, their thermodynamic behaviors and an effective single-band model. With the single-band picture, the superfluid density, Knight shift and the spin relaxation rate are discussed. We also make a comparative study between the honeycomb and the square lattices and find that the global phase diagram of these two systems is similar in spite of the topology of lattice.

The remainder of this paper is organized as follows.
In Sec.\ref{sec2}, t-t$^\prime$-J model on the honeycomb lattice is introduced and its mean-field formalism is derived. In Sec.\ref{sec3}, the antiferromagnetic normal state is detailed studied and the corresponding
Fermi surface topology is shown. Next, Sec.\ref{sec4} provides the results on the superconducting state, where the local density of state and the thermodynamic entropy are calculated. An effective single-band model for
superconducting state is also proposed in this section.
A brief comparison to the square lattice case is given in Sec.\ref{sec5}
Finally, we end this work with a brief conclusion in Sec.\ref{sec6}.

\section{The model and mean-field approximation} \label{sec2}
The t-t$^\prime$-J model on the honeycomb lattice is defined as follows,
\begin{eqnarray}
H&&=\hat{P}[-t\sum_{\langle ij\rangle\sigma}(a_{i\sigma}^{\dag}b_{j\sigma}+b_{i\sigma}^{\dag}a_{j\sigma})\nonumber \\
&&-t'\sum_{\langle\langle ij\rangle\rangle\sigma}(a_{i\sigma}^{\dag}a_{j\sigma}+b_{i\sigma}^{\dag}b_{j\sigma})]\hat{P}+J\sum_{\langle ij\rangle}\vec{S}_{i}\cdot\vec{S}_{j}.\label{eq1}
\end{eqnarray}
Here, as usual, the projection operator $\hat{P}=\prod_{i}(1-n_{i\uparrow}n_{i\downarrow})$ prohibits any double occupation on each site.
The $t$ ($t'$) term denotes nearest (next-nearest) neighbor hopping between different (the same) sublattice. The Heisenberg exchange term ($J$-term) considered here works for only nearest neighbor sites,
although more long-ranged spin-spin interaction (e.g. the next-nearest exchange term, which competes with the nearest one and leads to frustration effect) can be readily added.

Generically, the t-J like model is derived from more fundamental/microscopic Hubbard model in the strong coupling limit, where the double occupation is not allowed in the physical Hilbert space due to the large Hubbard U term.\cite{Lee2006}
However, the t-J model can also be an effective low-energy model from more complex model Hamiltonian, as well-known in the high temperature superconductivity of cuprate.\cite{Zhang1988}
For the present case with honeycomb lattice, the single-band Hubbard model at half-filling is detailed studied by many numerical simulations.\cite{Sorella1992,Meng,Sorella2012} The conclusion from those studies is that the ground-state is the usual antiferromagnetic state in the strong coupling limit while the usual Dirac metal state is stable in the weak coupling limit. The t-J model on the honeycomb lattice is discussed by RMFT\cite{Wu2013} and
GTPS\cite{Gu2013} approach as well. Interestingly, as shown in the GTPS calculation, there may exist a $d+id$ superconducting state, which coexists with collinear antiferromagnetic spin-density-wave (SDW) state at low doping.

Following the conventional slave-boson mean-field treatment,\cite{Yuan2004} the hopping term can be approximated as
\begin{eqnarray}
H_{0}&&=\sum_{k\sigma}[-t\delta f(k)f_{a k\sigma}^{\dag}f_{b k\sigma}+h.c.\nonumber \\
&&- (t'\delta\gamma(k)+\mu)(f_{a k\sigma}^{\dag}f_{a k\sigma}+f_{b k\sigma}^{\dag}f_{b k\sigma})],\label{eq2}
\end{eqnarray}
where the chemical potential $\mu$ is added with $a_{i\sigma}\simeq\sqrt{\delta}f_{a i\sigma}$ and $b_{i\sigma}\simeq\sqrt{\delta}f_{b i\sigma}$. ($\delta$ denotes the doping level deviated from half-filling and the total filling density is $n_{c}=2(1-\delta)$.) We have also defined $f(k)=e^{ik_{x}}+2e^{-ik_{x}/2}\cos(\sqrt{3}k_{y}/2)$ and $\gamma(k)=2\cos(\sqrt{3}k_{y})+4\cos(\sqrt{3}k_{y}/2)\cos(3k_{x}/2)$ and find that $\gamma(k)+3=|f(k)|^{2}$.
From the above expression, a noticeable effect of doping is to reduce the hopping energy, thus the interaction effect is more important in the present case than the unprojected free electrons.
For the particular half-filling case, the hopping term totally vanishes and the t-J model reduces to the simple Heisenberg model, which gives rise to an insulating ground-state with collinear antiferromagnetism.
Meanwhile, with the slave-boson representation $\vec{S}_{a (b)}=f_{a(b)}^{\dag}\frac{\vec{\sigma}}{2}f_{a(b)}$ , the Heisenberg exchange term can be approximated as
\begin{eqnarray}
&&S^{z}_{i}S^{z}_{j}\simeq m^{2}-\frac{m}{2}\sum_{\sigma}\sigma (f_{ai\sigma}^{\dag}f_{ai\sigma}-f_{bi\sigma}^{\dag}f_{bi\sigma}),\nonumber \\
&&\frac{1}{2}(S^{+}_{i}S^{-}_{j}+H.c.)\simeq \frac{-\chi}{2}\sum_{\sigma}[f_{ai\sigma}^{\dag}f_{bj\sigma}+H.c.]+\chi^{2}\label{eq3}
\end{eqnarray}
with the self-consistent staggered magnetization $m=(-1)^{i}<S_{i}^{z}>$ and valence-bond order $\chi=<f_{ai\sigma}^{\dag}f_{bj\sigma}>$. Note that, the staggered magnetization is not included in usual slave-boson treatment since one focuses on the possible nonmagnetic quantum spin liquids or non-Fermi liquids. As a matter of fact, such an antiferromagnetic SDW approximation is crucial for realistic applications such as
electron-doped cuprate,\cite{Kusko2002,Yuan2004,Xiang2009,Armitage2010} where the antiferromagnetism persists up to optimal doping and coexists with superconductivity around optimal doping. On the other hand, the candidate honeycomb insulating compound In$_{3}$Cu$_{2}$VO$_{9}$ seems to yield an antiferromagnetic ordered state at low temperature.\cite{Mydosh2008,Chen2012} With this real material in mind, it is reasonable to explicitly include the antiferromagnetism from the beginning. This is rather different from RMFT of Ref.\cite{Wu2013}, where only nonmagnetic states are considered.

Furthermore, if one considers the particle-particle channel or pairing instability, the
Heisenberg exchange term can be rewritten as
\begin{eqnarray}
\vec{S}_{i}\cdot\vec{S}_{j}=-\frac{1}{2}[(f_{ai\uparrow}^{\dag}f_{bj\downarrow}^{\dag}-f_{ai\downarrow}^{\dag}f_{bj\uparrow}^{\dag})(f_{bj\downarrow}f_{ai\uparrow}-f_{bj\uparrow}f_{ai\downarrow})]+\frac{1}{4}\nonumber
\end{eqnarray}
Then, defining the pairing order parameter $<f_{i\uparrow}^{\dag}f_{j\downarrow}^{\dag}-f_{i\downarrow}^{\dag}f_{j\uparrow}^{\dag}>=-\Delta_{ij}$,
we have
\begin{eqnarray}
\vec{S_{i}}\cdot\vec{S_{j}}\simeq\frac{1}{2}[\Delta_{ij}^{\star}(f_{i\uparrow}^{\dag}f_{j\downarrow}^{\dag}-f_{i\downarrow}^{\dag}f_{j\uparrow}^{\dag}) +H.c.+|\Delta_{ij}|^{2}].\label{eq4}
\end{eqnarray}
With different choice of $\Delta_{ij}$, one can detect possible paring instability based on the lattice symmetry of honeycomb lattice. Since the t-t$'$-J model possesses antiferromagnetic interaction built in,
in general, the singlet pairing is more relevant than the triplet pairing structure. Thus, the candidate pairing structure for the present case should be extended s-wave (uniform s-wave is suppressed due to no double-occupation condition), d-wave, and so on. Furthermore, according to the previous numerical\cite{Gu2013} and analytical\cite{Wu2013} analysis, the chiral $d+id$ pairing state is energetically more favored than any other pairing possibility. Therefore, we mainly focus on this $d+id$ pairing when superconducting states are involved. [We have checked that the $d+id$ pairing is energetically more favored than the extended s-wave in our mean-field calculation, which is
consistent with the phenomenological model calculation of Ref.\cite{Schaffer2007}.] A careful reader may wonder why extended s-wave state is not favored in t-t$'$-J model on honeycomb. Crudely, it is due to the elusive fermiology which means that for a generic single Fermi surface, the attraction in the extended-s wave channel is suppressed by the local Coulomb interaction.\cite{Coleman}[The local Coulomb interaction is not included in the present model for simplicity but it always appears in realistic situations.] Only when other compensating Fermi surface in regions with opposite sign structure appears, the Fermi surface average of the superconducting gap function will vanish, which permits a decoupling of the extended s-wave pairing from the repulsive Coulomb interaction. [This case appears in the superconductivity of iron-based compounds.\cite{Stewart2011}] In contrast, as a result of symmetry, the d-wave pairing decouples from the repulsive Coulomb interaction, thus it is more favored
and ubiquitous in model calculations and real materials.

\section{Antiferromagnetic normal state}\label{sec3}
In this section, the nonsuperconducting normal state of t-t$'$-J model on honeycomb lattice is inspected from the slave-boson mean-field treatment. As what has been explained in last section, the antiferromagnetism is explicitly considered from the beginning and the resulting mean-field Hamiltonian reads [using Eqs.(\ref{eq2}) and (\ref{eq3})]
\begin{eqnarray}
H_{AFM}&&=\sum_{k\sigma}[(-t\delta f(k)-\frac{J\chi}{2})f_{a k\sigma}^{\dag}f_{b k\sigma}+h.c.\nonumber\\
&&+(-t'\delta\gamma(k)-\mu)(f_{a k\sigma}^{\dag}f_{a k\sigma}+f_{b k\sigma}^{\dag}f_{b k\sigma})\nonumber\\
&&-\frac{3Jm\sigma}{2}(f_{a k\sigma}^{\dag}f_{a k\sigma}-f_{b k\sigma}^{\dag}f_{b k\sigma})] +3J(\chi^{2}+m^{2}).\label{eq5}
\end{eqnarray}
This Hamiltonian has quasiparticle spectrum
\begin{eqnarray}
&&\xi_{k}^{\pm}=\pm E_{0k}-t'\delta\gamma(k)-\mu,\label{eq6}
\end{eqnarray}
where we have defined $E_{0k}=\sqrt{(t\delta+\frac{J\chi}{2})^{2}|f(k)|^{2}+\frac{9}{4}(Jm)^{2}}$. It is noted that an antiferromagnetic gap $\frac{3}{2}Jm$ is opened in $E_{0k}$. For a half-filling system with $t'=0$, the whole system is gapped and this corresponds to an antiferromagnetic Mott insulator but not a Slater insulator since the gap of the latter one is formed in terms of band-filling.

The free energy reads
\begin{eqnarray}
F=3J(\chi^{2}+m^{2})-2T\sum_{k,\alpha=\pm}\ln(1+e^{-\beta \xi_{k}^{\alpha}}),\nonumber
\end{eqnarray}
and the resulting mean-field self-consistent equations can be obtained by
\begin{equation}
\frac{\partial F}{\partial m^{2}}=0, \,\, \frac{\partial F}{\partial \chi}=0, \nonumber
\end{equation}
which read
\begin{eqnarray}
&& 1=\frac{3}{4}J\sum_{k}\frac{f_{F}(\xi_{k}^{-})-f_{F}(\xi_{k}^{+})}{E_{0k}}.\label{eq7}\\
&& 6\chi=(t\delta+\frac{J\chi}{2})\sum_{k}|f(k)|^{2}\frac{f_{F}(\xi_{k}^{-})-f_{F}(\xi_{k}^{+})}{E_{0k}}.\label{eq8}
\end{eqnarray}
Besides, the chemical potential $\mu$ is determined by $n_{c}=2(1-\delta)=-\frac{\partial F}{\partial \mu}$, which gives the last equation
\begin{eqnarray}
2(1-\delta)=2\sum_{k}[f_{F}(\xi_{k}^{-})+f_{F}(\xi_{k}^{+})].\label{eq9}
\end{eqnarray}

\begin{figure}
\includegraphics[width=0.8\columnwidth]{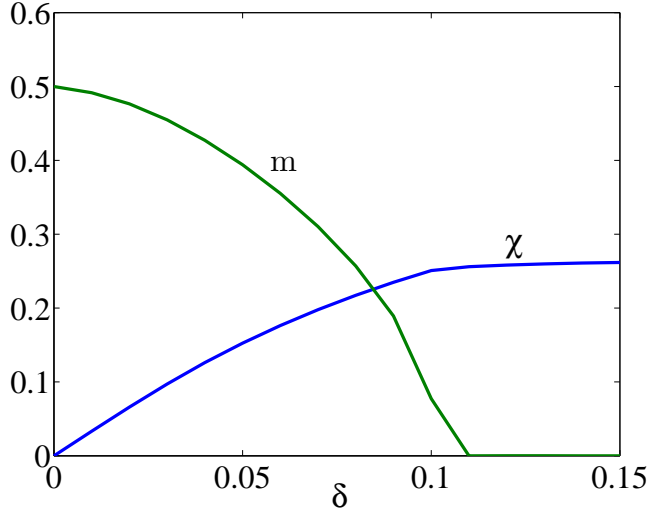}
\caption{\label{fig:1} The magnetization $m$ and valence-bond order $\chi$ as functions of doping in the antiferromagnetic normal state.}
\end{figure}

With Eqs. (\ref{eq7}) - (\ref{eq9}), the mean-field parameters $m$ and $\chi$ are calculated with the fixed temperature and doping $\delta$. In the present paper, we use
$t=-1,t=0.1,J=0.3$ and $T=0.01$.[$T=0.01$ is sufficient to get the correct result for the ground-state and we have checked that the lower temperature $T=0.001$ does not change the results obtained at $T=0.01$.]
In Fig.\ref{fig:1}, the mean-field parameters magnetization $m$ and valence-bond order $\chi$ are shown in the antiferromagnetic normal state. It is clear to see that the antiferromagnetic SDW state persists up to
$\delta=0.11$ and vanishes for larger doping level. Surprisingly, the position of the vanishing point of magnetization agrees fairly well with density-matrix-renormaliztion-group and GTPS calculation,\cite{Gu2013} which suggests that the qualitative or even quantitative physics is correctly captured by the present straightforward mean-field approximation. Meanwhile, the valence-bond order $\chi$, which denotes the kinetic energy gained by resonant spin-singlet exchange, increases upon doping and reach it maximal value after $\delta=0.11$. Now, let us try to understand the mentioned doping evolution more intuitively. At half-filling, the kinetic energy is spoiled by
the no double occupation condition and the system organizes itself to form the bipartite antiferromagnetism to optimize the energy of ground-state. Then, upon doping, the Neel antiferromagnetic state is frustrated by the appearance of doped holes and the kinetic energy is strengthened by more available empty sites. When a critical doping level approaches, the antiferromagnetic background is killed and the kinetic energy finally wins.

\begin{figure}
\includegraphics[width=0.8\columnwidth]{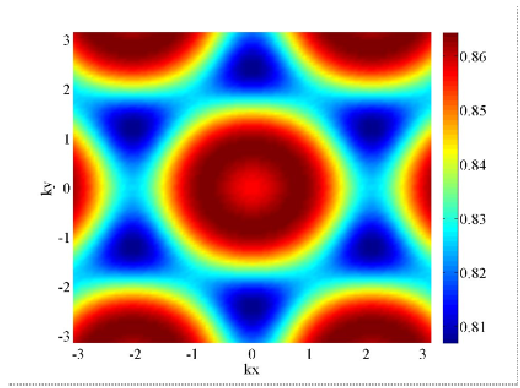}
\caption{\label{fig:2} The calculated ARPES intensity of the antiferromagnetic normal state at doping level $\delta=0.02$.}
\end{figure}

\begin{figure}
\includegraphics[width=0.8\columnwidth]{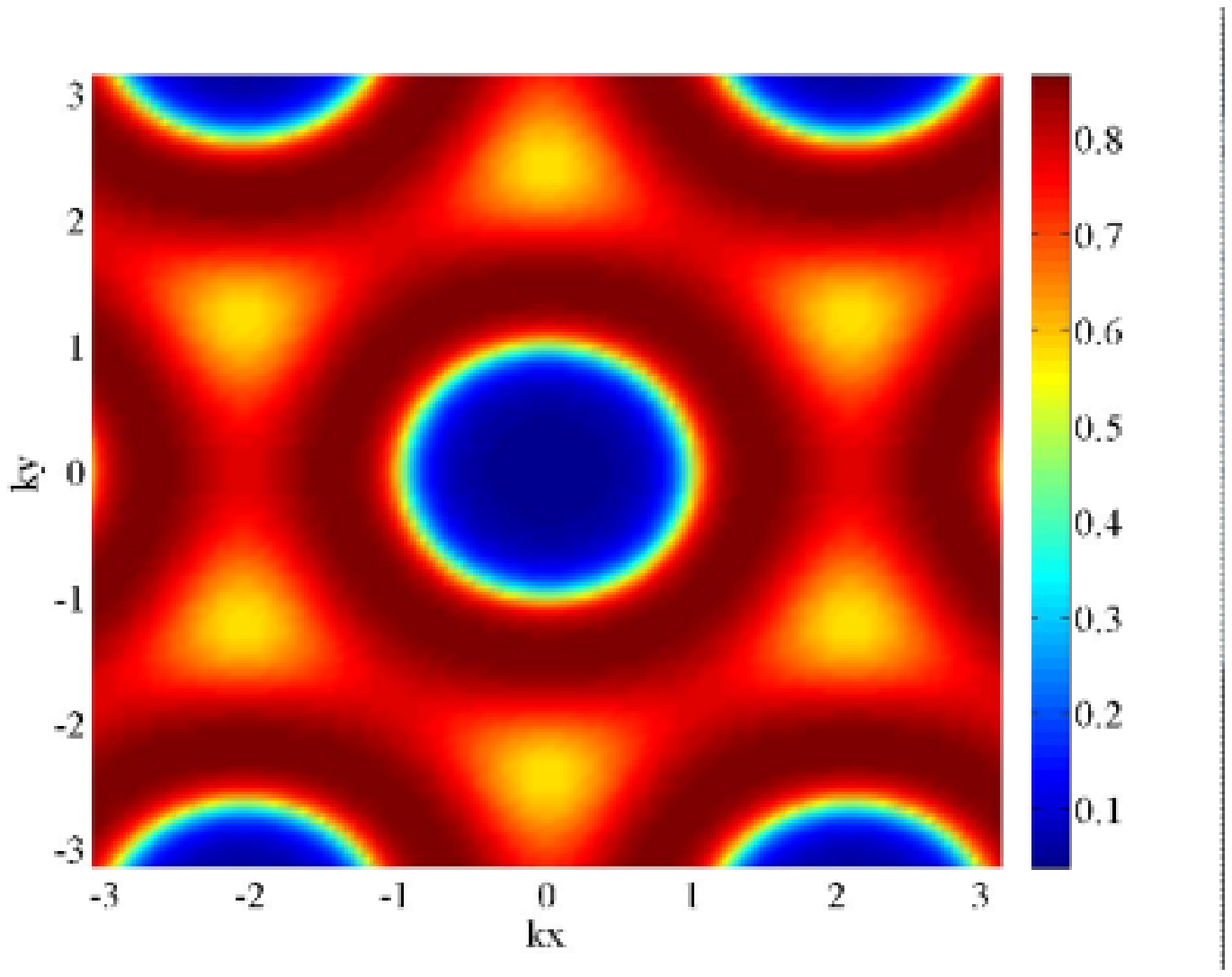}
\caption{\label{fig:3} The calculated ARPES intensity of the antiferromagnetic normal state at doping level $\delta=0.05$}
\end{figure}

\begin{figure}
\includegraphics[width=0.8\columnwidth]{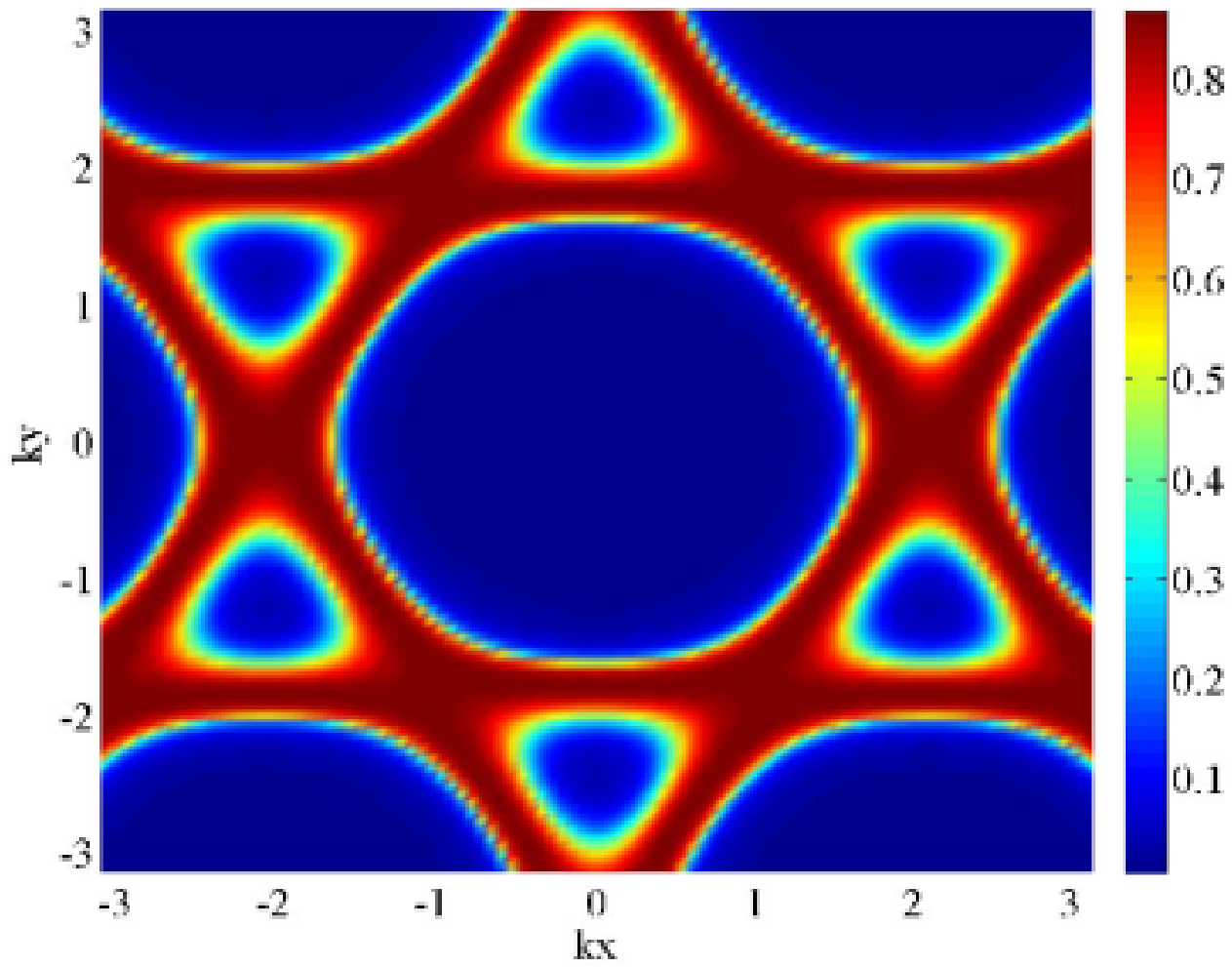}
\caption{\label{fig:4} The calculated ARPES intensity of the antiferromagnetic normal state at doping level $\delta=0.10$}
\end{figure}

Experimentally, the angle-resolved photoemission spectroscopy (ARPES) is often used to detect single-particle information and the Fermi surface topology. Here, the corresponding
ARPES intensity of the antiferromagnetic normal state is shown in Figs.\ref{fig:2},\ref{fig:3} and \ref{fig:4}, which is defined as
\begin{eqnarray}
I(k)&&=\int_{a}^{b} d\omega A(k,\omega)=\int_{a}^{b} d\omega[ -\frac{1}{\pi}TrG(k,\omega)]\nonumber\\
&&\approx \delta\int_{a}^{b} d\omega [\frac{\Gamma/\pi}{(\omega-\xi_{k}^{+})^{2}
+(\Gamma/\pi)^{2}}+\frac{\Gamma/\pi}{(\omega-\xi_{k}^{-})^{2}+(\Gamma/\pi)^{2}}]\nonumber
\end{eqnarray}
with $a=-b=0.1$ are the integrated regime of usual experiments, and we take $\Gamma=0.01$.
The ARPES intensity shows clearly that a single-band with large Fermi surface (basically formed by $\xi_{k}^{-}$) is active and this is crucial for the domination of d-wave pairing over the extended s-wave one.

\section{Superconducting state with antiferromagnetism }\label{sec4}
In this section, we discuss the mean-field solution of possible superconducting (SC) state with antiferromagnetism. Here, we do not immerse into the elusive issue of pairing mechanism but only mention that the antiferromagnetic Heisenberg exchange term is able to induce the superconducting pairing based on the resonance-valence-bond (RVB) picture\cite{Lee2006,Anderson2004} or antiferromagnetic spin fluctuation framework.\cite{Miyake1986,Scalapino1986,Emery1986} We should emphasize that the use of t-J-like model does not always imply that its pairing mechanism is RVB. In some sense, one may use the renormalized t-J-like model\cite{Anderson2004}
and analyze its pairing instability in terms of perturbative renormalization group,\cite{Nandkishore2014} which could be a starting point for spin fluctuation analysis.\cite{Scalapino2012} In addition, we note that functional renormalization group (fRG) calculations are made for a general Hamiltonian on the honeycomb lattice.\cite{Honerkamp2008,Wu2013} They find that the system appears to flow toward a d+id superconducting state as the temperature is lowered.
The driven force of superconducting pairing seems to be the antiferromagnetic spin fluctuation in this framework, but for the present t-J-like model it is not clear whether this mechanism really works.
We should remind the reader that due to the notorious projection operator, the functional renormalization group method cannot directly attack the t-J-like model, thus it is still unknown whether antiferromagnetic spin fluctuation mechanism is responsible for the d-wave pairing in the strongly coupled t-J-like model. However, if one believes that result of the weak and intermediate coupling regime is able to smoothly evolve into strong coupling regime, the fRG might have given the solution in such systems. [To our knowledge, no functional renormalization group calculations have been made in t-J-like models.]

Using Eqs. (\ref{eq4}) and (\ref{eq5}), the mean-field Hamiltonian for the superconducting state reads
\begin{eqnarray}
H_{SCA}&&=\sum_{k\sigma}[(-t\delta f(k)-\frac{J\chi}{2})f_{a k\sigma}^{\dag}f_{b k\sigma}+h.c.\nonumber\\
&&-(t'\delta\gamma(k)+\mu)(f_{a k\sigma}^{\dag}f_{a k\sigma}+f_{b k\sigma}^{\dag}f_{b k\sigma})\nonumber\\
&&-\frac{3Jm\sigma}{2}(f_{a k\sigma}^{\dag}f_{a k\sigma}-f_{b k\sigma}^{\dag}f_{b k\sigma})] \nonumber\\
&&+\frac{J}{2}\sum_{k}[\Delta_{k}f_{a -k\downarrow}^{\dag}f_{b k\uparrow}^{\dag}+\Delta_{k}f_{b -k\downarrow}^{\dag}f_{a k\uparrow}^{\dag}+h.c.]\nonumber\\
&&+3J(\chi^{2}+m^{2}+\frac{1}{2}\Delta^{2}).\label{eq10}
\end{eqnarray}
Here, the $d+id$-wave pairing gap function $\Delta_{k}=\Delta\Gamma(k)$ with $\Gamma(k)=e^{ik_{x}}+e^{-ik_{x}/2}(\sqrt{3}\sin(\sqrt{3}k_{y}/2)-\cos(\sqrt{3}k_{y}/2))$.\cite{Wu2013}
In addition, if one is interested in the extended s-wave pairing, the corresponding pairing gap function reads as $\Delta_{k}=\Delta f(k)$ with $f(k)=e^{ik_{x}}+2e^{-ik_{x}/2}\cos(\sqrt{3}k_{y}/2)$.\cite{Kotov2012}
Note that the true superconducting pairing function should be $\Delta_{k}^{SC}=\delta \Delta_{k}$ with the strong doping dependence $\delta$. Therefore, a superconducting dome may form if $\Delta$ decreases as
doping increases.

The corresponding mean-field equations are readily derived as
\begin{eqnarray}
&&1=\frac{3J}{8}\sum_{k}\frac{1}{E_{0k}}(\frac{\xi_{k}^{+}}{E_{k}^{+}}\tanh\frac{E_{k}^{+}}{2T}-\frac{\xi_{k}^{-}}{E_{k}^{-}}\tanh\frac{E_{k}^{-}}{2T}),\nonumber\\
&&6\chi=\frac{1}{2}(t\delta+\frac{J\chi}{2})\sum_{k}\frac{|f(k)|^{2}}{E_{0k}}(\frac{\xi_{k}^{+}}{E_{k}^{+}}\tanh\frac{E_{k}^{+}}{2T}-\frac{\xi_{k}^{-}}{E_{k}^{-}}\tanh\frac{E_{k}^{-}}{2T}),\nonumber\\
&&\frac{3}{2}=\frac{J}{4}\sum_{k}|\Gamma(k)|^{2}(\frac{1}{2E_{k}^{+}}\tanh\frac{E_{k}^{+}}{2T}+\frac{1}{2E_{k}^{-}}\tanh\frac{E_{k}^{-}}{2T}),\nonumber\\
&&2(1-\delta)=\sum_{k}(2-\frac{\xi_{k}^{+}}{E_{k}^{+}}\tanh\frac{E_{k}^{+}}{2T}-\frac{\xi_{k}^{-}}{E_{k}^{-}}\tanh\frac{E_{k}^{-}}{2T}).\nonumber
\end{eqnarray}
Here, the SC quasiparticle energy is defined as $E_{k}^{\pm}=\sqrt{(\xi_{k}^{\pm})^{2}+\frac{J^{2}}{4}|\Delta_{k}|^{2}}$.

\begin{figure}
\includegraphics[width=0.9\columnwidth]{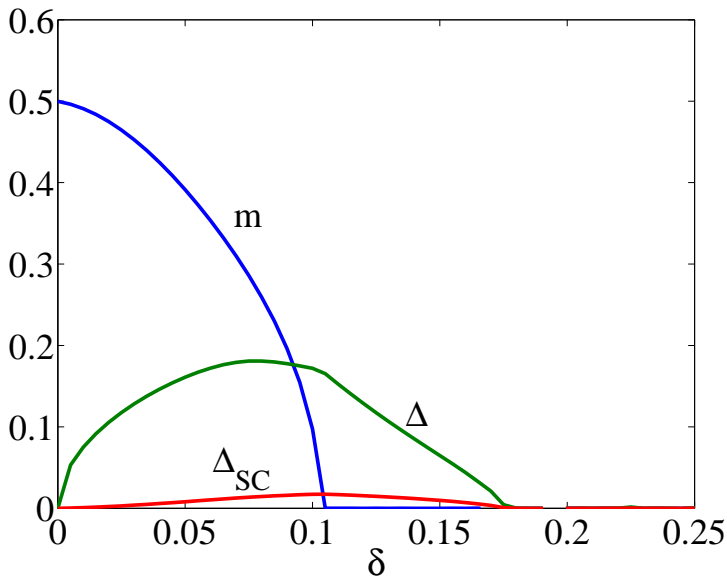}
\caption{\label{fig:5} The doping evolution of the mean-field parameters magnetization $m$, pairing order $\Delta$ and true SC pairing order $\Delta_{SC}$ are shown in superconducting state.}
\end{figure}

The mean-field parameters $m$, $\Delta$ and true SC pairing order $\Delta_{SC}$ are shown in Fig.\ref{fig:5}.
It can be seen that there exists a large coexistent regime for antiferromagnetic SDW and superconducting states up to doping level $\delta\simeq0.11$. Interestingly,
this value of doping is exactly identical to the one in the pure antiferromagnetic SDW phase discussed in last section, which suggests
that the two symmetry-breaking phases are not competing so strongly at least in the present mean-field level. However, we should emphasize that if fluctuation beyond mean-field is included as shown in the typical fRG calculation,\cite{Friederich2011} antiferromagnetic and superconducting order will show a tendency of mutual exclusion, which indicates a strong competition in contrast to the mean-field result. Another interesting point is that the maximal SC pairing order appears when the
antiferromagnetic SDW just vanishes. In some sense, this feature is due to the interplay of pairing and antiferromagnetic order, but the detail of this issue is still unknown to our knowledge. [The position where the d-wave order parameter vanishes seems to deviate from the one given in Ref.\cite{Wu2013}, which is due to a different choice of the exchange coupling parameter and the inclusion of $t'$. The choice in the preset work has the benefit to link with numerical calculations in GTPS.]

In addition, when comparing the mean-field result with the numerical simulation of GTPS,\cite{Gu2013} to our surprised, the coexistent regime found in our present paper agrees rather well with the coexisting regime $0<\delta<0.1$ obtained by GTPS. This implies that the simple and physically transparent mean-field theory has captured the basic feature of the t-t$'$-J model on honeycomb lattice

\subsection{The local density of state of superconducting state}
The local density of state (LDOS) of the superconducting state, which can be readily measured by scanning tunneling microscopy\cite{Fischer2007} or point-contact spectroscopy experiments,
has a simple form
\begin{eqnarray}
N(\omega)=\sum_{k,\alpha = \pm}[\delta(\omega-E_{k}^{\alpha})+\delta(\omega+E_{k}^{\alpha})].\nonumber
\end{eqnarray}

\begin{figure}
\includegraphics[width=0.8\columnwidth]{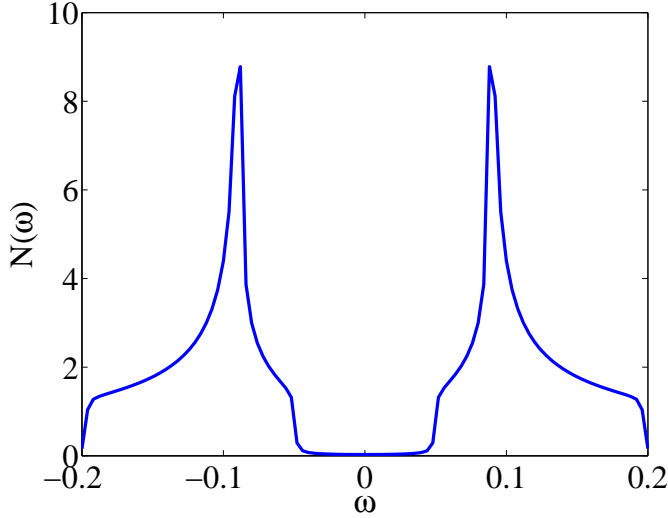}
\caption{\label{fig:6} The local density of state in superconducting state at doping $\delta=0.05$.}
\end{figure}

\begin{figure}
\includegraphics[width=0.8\columnwidth]{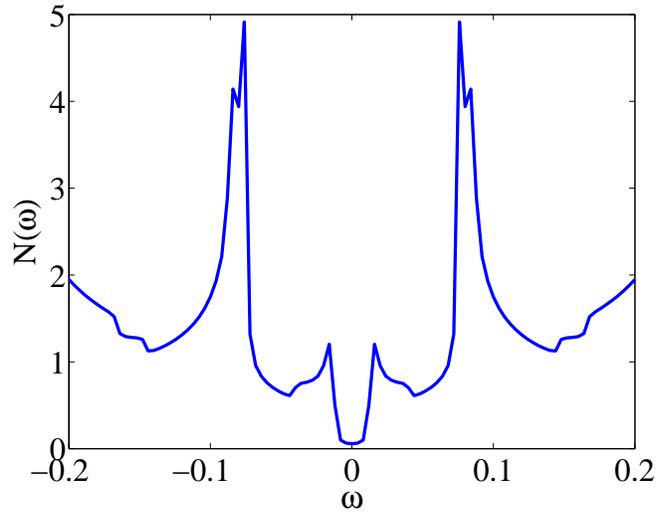}
\caption{\label{fig:7} The local density of state in superconducting state at doping $\delta=0.10$.}
\end{figure}
In Figs. \ref{fig:6} and \ref{fig:7}, the typical LDOS is shown and the full SC gap is clearly shown as indicated by the BCS coherent peak, which is an explicit signature of the generic $d+id$ superconducting pairing state.
Note that the large peak at high energy ($\omega \sim 0.09$) just reflects the van Hove singularity point but not the familiar BCS coherent peak.

\subsection{The thermodynamic entropy of superconducting state}
Another useful physical observable of SC phase is the thermodynamic entropy, whose expression reads
\begin{eqnarray}
S=\frac{2}{T}\sum_{k,\alpha=\pm}[E_{k}^{\alpha}f_{F}(E_{k}^{\alpha})+T\ln(1+e^{-\beta E_{k}^{\alpha}})].\nonumber
\end{eqnarray}
\begin{figure}
\includegraphics[width=0.8\columnwidth]{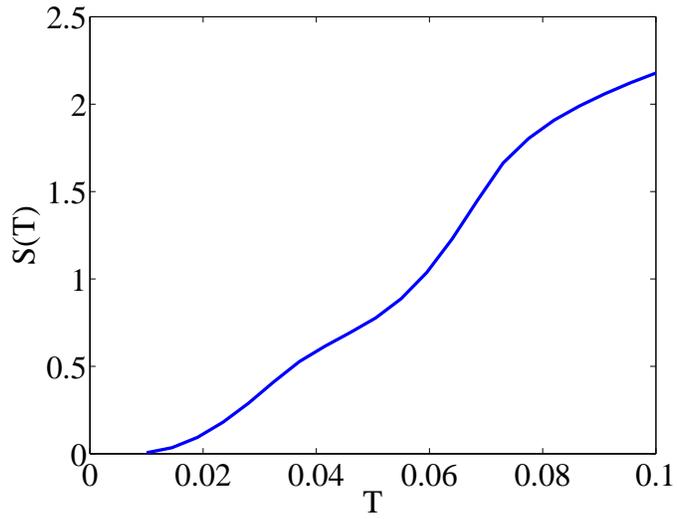}
\caption{\label{fig:8} The thermodynamic entropy in superconducting state at doping $\delta=0.05$.}
\end{figure}

\begin{figure}
\includegraphics[width=0.8\columnwidth]{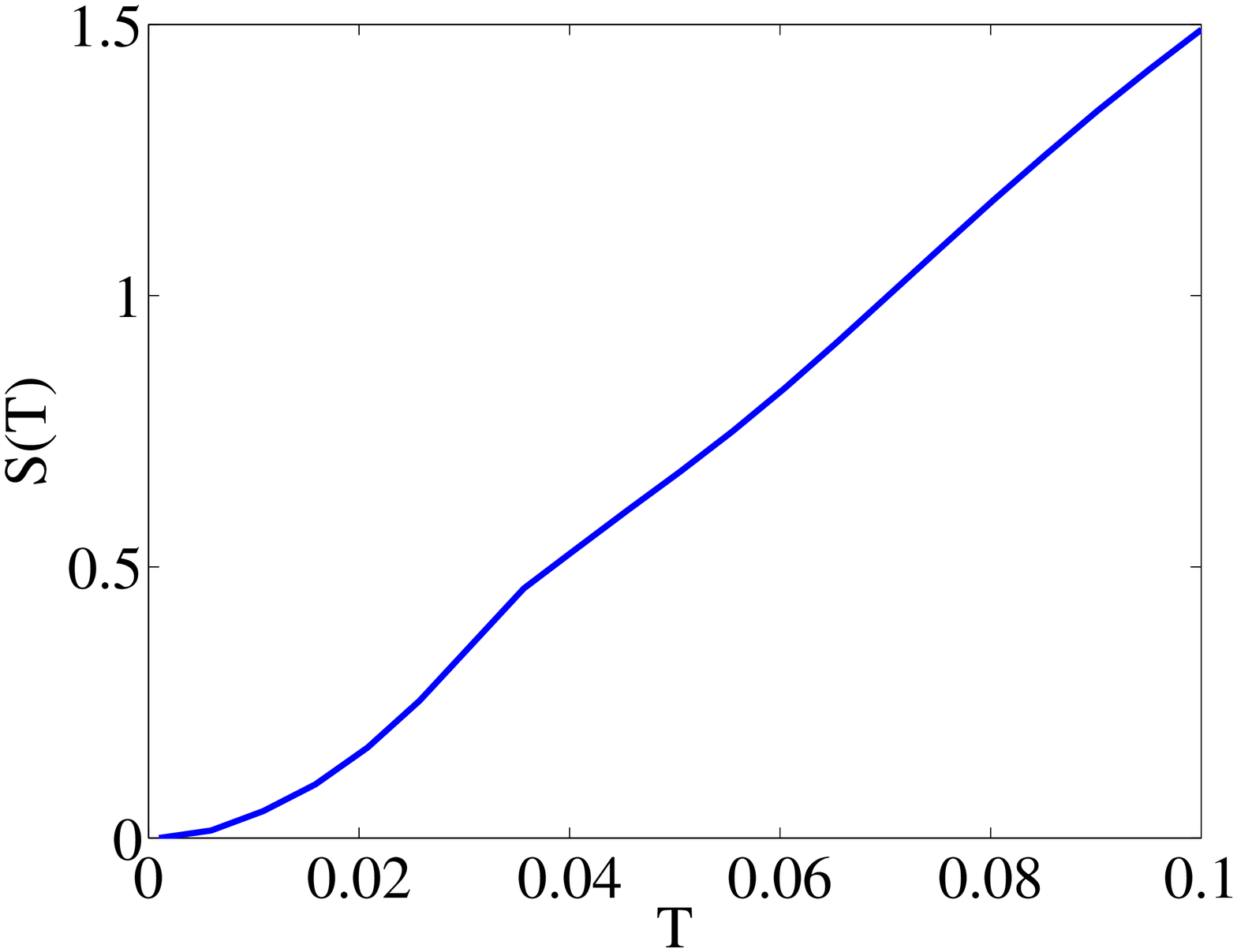}
\caption{\label{fig:9} The thermodynamic entropy in superconducting state at doping $\delta=0.10$.}
\end{figure}

From Figs.\ref{fig:8} and \ref{fig:9}, the entropy at low temperature is suppressed, which also consists with the gapped $d+id$ pairing state. (A fitting with $S\sim T^{2}$ is not good, thus the gapless (nodal) SC state is excluded.)

\subsection{An effective single-band picture for superconducting state}
As noted in last section, the active band is only the $\xi_{k}^{-}$ band and it is helpful to develop an effective single-band picture to understand
the basic properties of superconducting state. First, the antiferromagnetic mean-field Hamiltonian $H_{AFM}$ can be diagonalized as
\begin{eqnarray}
H_{AFM}=\sum_{k\sigma}[\xi_{k}^{+}A_{k\sigma}^{\dag}A_{k\sigma}+\xi_{k}^{-}B_{k\sigma}^{\dag}B_{k\sigma}]\nonumber
\end{eqnarray}
by using the transformation relations between original fermions and antiferromagnetic quasiparticle $f_{k\sigma}^{a}=\frac{1}{\sqrt{2}}[(\alpha_{k}-\sigma\beta_{k})A_{k\sigma}+(\alpha_{k}+\sigma\beta_{k})B_{k\sigma}]$ and
$f_{k\sigma}^{b}=\frac{e^{-i\theta_{k}}}{\sqrt{2}}[(\alpha_{k}+\sigma\beta_{k})A_{k\sigma}+(-\alpha_{k}+\sigma\beta_{k})B_{k\sigma}]$. Here, we have defined $\alpha_{k}^{2}=1-\beta_{k}^{2}=\frac{1}{2}(1+\frac{\bar{\varepsilon}_{k}}{E_{0k}})$ with $\bar{\varepsilon}_{k}=\sqrt{(t\delta+J\chi/2)^{2}|f(k)|^{2}}$ and $\theta_{k}=\arctan(\frac{Im f(k)}{Re f(k)})$.

Then, adding the pairing term and neglecting the contribution of $\xi_{k}^{+}$, namely the $A_{k\sigma}$ quasiparticles, we have
the expected effective single-band superconducting  model
\begin{eqnarray}
H=\sum_{k\sigma}[\xi_{k}^{-}B_{k\sigma}^{\dag}B_{k\sigma}]+\sum_{k}[\tilde{\Delta}_{k}B_{k\uparrow}^{\dag}B_{-k\downarrow}^{\dag}+h.c.]\nonumber,
\end{eqnarray}
where the effective gap function $\tilde{\Delta}_{k}=\frac{J}{2}\Delta_{k}(-\cos\theta_{k}-2i\alpha_{k}\beta_{k}\sin\theta_{k})$. Basically, the nodal properties of $\tilde{\Delta}_{k}$ is similar to $\Delta_{k}$, thus
the $d+id$ feature of $\Delta_{k}$ is preserved in the present simplified model. Utilizing this single-band model, many physical observable can be readily calculated or argued.\cite{Xiang}

\subsection{Remark on other observables}
\begin{figure}
\includegraphics[width=0.80\columnwidth]{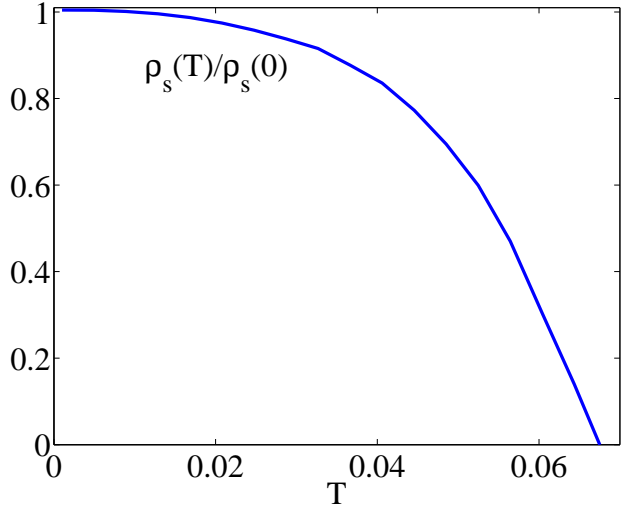}
\caption{\label{fig:10} The normalized superfluid density $\rho_{s}(T)/\rho_{s}(0)$ versus temperature T.}
\end{figure}

The superfluid density, which directly detects the quasiparticle of SC phase, is expected to
have similar behavior as the usual uniform s-wave case since the low energy excitations are all gaped. [The low temperature part of superfluid density is determined by the DOS of SC quasiparticle \cite{Xiang} and the
gapped $d+id$-wave leads to the gapped DOS as shown in previous calculation.]
We also note that in contrast to the case in the electron-doped cuprate,\cite{Luo2005,Das2007} the single-band superfluid density formula can be used for the present case as
what has been discussed in last subsection. Using the effective single-band superfluid density formula $\rho_{s}(T)=\sum_{k}\left[-\frac{\partial^{2} \xi_{k}^{-}}{\partial k_{\mu}^{2}}\frac{\xi_{k}^{-}}{E_{k}^{-}}\tanh(\frac{E_{k}^{-}}{2T})+2(\frac{\partial \xi_{k}^{-}}{\partial k_{\mu}})^{2}\frac{\partial f_{F}(E_{k}^{-})}{\partial E_{k}^{-}}\right]$, \cite{Xiang} we have plotted its typical result in Fig.\ref{fig:10} with doping $\delta=0.07$, which is consistent with the exponential behavior at low temperature.

The temperature dependent magnetization should behave as the superfluid density, which shows exponential behavior at low temperature.
Moreover, with the single-band picture, the temperature dependence of the Knight shift ($K_{s}$) and the spin relaxation rate
($\frac{1}{T_{1}}$) are\cite{Zhou2008}
\begin{eqnarray}
\{K_{s},\frac{1}{T_{1}}\}\propto \int d\omega \{1,TN(\omega)\}N(\omega)\frac{\partial f_{F}(\omega)}{\partial\omega}\nonumber.
\end{eqnarray}
Since the DOS ($N(\omega)$) is gapped at low energy, we expect that both $K_{s}$ and $\frac{1}{T_{1}}$ have exponential decays at low temperatures, which may be detected in future experiments.

\section{Comparison to the square lattice case}\label{sec5}
\begin{figure}
\includegraphics[width=0.8\columnwidth]{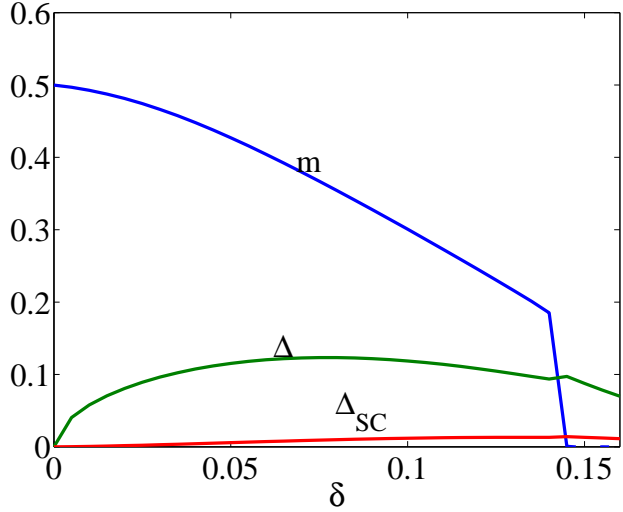}
\caption{\label{fig:11} The doping evolution of the mean-field parameters magnetization $m$ and pairing order $\Delta$ and true SC pairing order $\Delta_{SC}$ on the square lattice.}
\end{figure}

In the main text, we have studied the basic feature of t-t$'$-J model on the honeycomb lattice. Here, it is interesting to compare it to the more familiar case on the square lattice.
Because the antiferromagnetism plays a major role in our present model, we should compare it to the t-t$'$-J model for electron-doped cuprate, where the AF long-ranged order persists up to $\delta\sim0.15$ and coexists with
SC phase. For the antiferromagnetic normal state, the doping evolution of magnetization is similar to that in Ref.\cite{Yuan2004}. We note that the AF order vanishes at lower doping level in our case, which is due to the low nearest-neighbor number. However, the Fermi surface topology is rather different since the one on the square lattice shows two-band behavior while only one band is active in our case. As for the superconducting state, using the same method in Ref.\cite{Yuan2004}, we obtain the doping evolution of the mean-field parameters magnetization $m$ and pairing order $\Delta$ and true SC pairing order $\Delta_{SC}$ on the square lattice case in Fig.\ref{fig:11}.
We can see that the doping evolution of the mean-field parameters is similar when compared to the honeycomb lattice case of Fig.\ref{fig:5} but it should be emphasized that the SC pairing on the square lattice is
the usual $d_{x^{2}-y^{2}}$-wave while our case is $d+id$-wave. The differences on physical quantities like LDOS or entropy are obvious due to the gaplessness of $d_{x^{2}-y^{2}}$-wave.
The topological properties of $d+id$ superconducting state can also be understood from the single-band model and we do not discuss it but refer the Reader to Ref.\cite{Zhou2008}.

\section{Conclusion}\label{sec6}
In this paper we study systematically the t-t$'$-J model on honeycomb lattice by the slave-boson mean-field method. It is found that the antiferromagnetism as a function of doping is consistent with the existing numerical calculation in the normal state. When the superconducting instability is considered, the superconductivity and antiferromagnetism can coexist in a broad doping regime, which is again in good agreement with the numerical calculation.
These results indicate that the slave-boson mean-field theory is a simple but reliable method in treating such strongly correlated systems, specifically, in the presence of antiferromagnetic long-range order. We also further explore the local density of states, its thermodynamic properties of the superconducting state and the transport behaviors like the superfluid density, which are useful to further experimental study on this honeycomb compound In$_{3}$Cu$_{2}$VO$_{9}$, specially, the possible superconductivity by introducing carriers. Our work is also helpful in understanding of the unconventional superconductivity on general two-dimensional correlated electron systems.

\section*{Acknowledgments}
The work was supported partly by NSFC, PCSIRT (Grant No. IRT1251), and the national program for basic research of China.

\section*{References}

\bibliography{mybibfile}

\end{document}